\documentclass[proceedings]{JHEP}
\def \lta{\mathrel{\vcenter{\hbox{$<$}\nointerlineskip\hbox{$\sim$}}}}

\newcommand{\beq}{\begin{equation}}
\newcommand{\eeq}{\end{equation}}
\newcommand{\beqs}{\begin{eqnarray}}
\newcommand{\eeqs}{\end{eqnarray}}
\newcommand{\dd}{{\mathrm d}}
\newcommand{\ee}{{\mathrm e}}

\newcommand{\de}{\partial}
\conference{Quantum aspects of gauge theories, supersymmetry and unification.}
\title{AdS/CFT Correspondence and Type 0 String Theory\footnote{Preprint SISSA
17/2000/EP}}
\author{Dario Martelli\\
	SISSA, Via Beirut 2-4 Trieste 34014\\
        INFN Sezione di Trieste, Italy\\
        dmartell@sissa.it}
\abstract{We review some applications of Type 0 string theory
in the context of the AdS/CFT correspondence.}

\begin{document}
Most of the success of the AdS/CFT correspondence is so far devoted to those
situations where at least some fraction of supersymmetry is preserved. The 
case of Type IIB supergravity in AdS$_5\times$S$^5$ is a typical example
\cite{Maldacena:1998re}, where
a detailed mapping between quantities living in the two sides of the
correspondence can be exploited.
However, an important issue to address is how to embed the physically relevant
nonsupersymmetric gauge theories in the correspondence, eventually recovering
asymptotic freedom and confinement.

There are different proposals for giving a holographic 
description of nonsupersymmetric gauge theories, which basically deal with  
possible mechanisms for breaking supersymmetry.
After recalling some of them, we will 
discuss some consequences of the approach proposed by Polyakov in
\cite{Polyakov:a}\footnote{See \cite{Polyakov:2000ti} for recent developments.} 
that consists in considering a nonsupersymmetric
string theory with non-chiral GSO projection, known as Type 0   
\cite{Dixon,Seiberg}. 

Before going to the main subject, we will briefly recall some of the 
alternative 
approaches presented in the literature.
  
It was pointed out by Witten \cite{Witten:1998qj,Witten:1998zw} that
the AdS/CFT correspondence 
can be formulated at {\it finite temperature}. In doing so one identifies
the Hawking temperature $T_H$ of the supergravity solution with that of the 
field theory in a thermal bath.
The antiperiodic boundary conditions for fermions along the compactified (time)
direction break supersymmetry and give them masses $m_f\sim T_H$ at tree level.
This leads also spin 0 bosons to acquire nonzero masses at one loop and
spoils conformal symmetry because cancellations in the $\beta$-function
do not occur any more. At large distances ($L\gg R_{comp}\sim\frac{1}{T_H}$)
the infrared effective theory is then expected to be pure YM in one 
lower dimension.

This approach captures some of the expected qualitative features of the quantum
field theory, as a {\it confining} behavior of the Wilson loop (basically
due to the presence of a horizon in the metric) and a {\it mass-gap} in the
glue-ball spectrum. Moreover it provides a Lorentz invariant regularization
scheme, as opposed for instance to the lattice regularization.
Nonetheless it has some drawbacks: There is still coupling
with the physics in higher dimension, as both the masses of glue-balls and
of fermions are of the same order of the Hawking temperature. 
In addition, the $p$-dimensional 
't Hooft coupling obeys

\begin{center}
$\lambda_p\sim \lambda_{p+1}T_H$
\end{center}
    
\noindent
where $T_H$ acts as UV cutoff. As the cutoff is removed ($T_H\to \infty $),
$\lambda_{p+1}$ should approach zero, but this is opposite to the regime
in which the supergravity approximation applies\footnote{In this respect the
situation is similar to the lattice approach, where computations are possible
at strong coupling.}.

Among various means to break supersymmetry there are the {\it deformations} of 
supersymmetric solutions. One searches for
(domain wall) solutions interpolating between some AdS vacuum with 
${\cal N}$ supersymmetries and a different vacuum that can be AdS or not, and has 
${\cal N}'<{\cal N}.$\footnote{See for instance the talks related to
\cite{Petrini:2000qa} in this
conference.} In the field theory side this is interpreted as turning on
some relevant operators that drive the initial theory to an
effective IR theory via RG flow of the
couplings. Later we will mention
an application of this construction in the context of Type 0 theory.

Another way to break supersymmetry is that of considering {\it orbifolds} theories,
or {\it branes at singularities}. These two methods have been pursued extensively 
in the literature and here we will not comment them further on.

\section{Type 0 Strings and Gravity}
In this section we summarize some basic features of Type 0 string theories
\cite{Dixon,Seiberg} and their gravity effective actions \cite{Klebanov:a}.

Type 0 string theories are purely bosonic strings with modular invariant
partition function and four sectors whose low laying fields 
are here summarized\footnote{The two entries in parenthesis refer
respectively to left and right movers and the signs correspond to the
choice of GSO projection. The upper (lower) signs in the RR sectors
define Type 0A (0B) theory.}

\begin{center}
\begin{tabular}{c|cc}
 & $(NS+,NS+)$ & $(NS-,NS-)$ \\
\hline 
{\bf 0A/0B} & $\Phi ,B_{\mu\nu} ,g_{\mu\nu}$ & $T$ \\
\end{tabular}
\vskip .5cm
\begin{tabular}{c|cc}
 & $(R+,R\mp)$ & $(R-,R\pm)$ \\
\hline 
{\bf 0A} &
$A^{(1)}_{\mu} A^{(1)}_{\mu\nu\rho}$ &
$A^{(2)}_{\mu} A^{(2)}_{\mu\nu\rho}$ \\
{\bf 0B} &  
$A^{(1)} A^{(1)}_{\mu\nu} A^{(+)}_{\mu\nu\rho\sigma}$ & 
$A^{(2)} A^{(2)}_{\mu\nu} A^{(-)}_{\mu\nu\rho\sigma}$ \\
\end{tabular}
\end{center}
There is a corresponding doubled set of D-branes coupling to RR-fields.

As pointed out in \cite{Polyakov:a} there are no open string tachyons
on the world-volume of these branes, while there is a perturbative closed
string tachyon that renders the Minkowski vacuum unstable. 
Nevertheless one should 
regard this as an indication that ten dimensional flat background is not 
stable, while there should exist other vacua in which the theory makes sense.
AdS space seems in this respect a good candidate in that it allows for 
tachyonic modes.

${\cal N}=(1,1)$ supersymmetry on the world-sheet makes these theories similar
in some 
respect to supersymmetric Type II. For example all tree level correlators 
of vertex 
operators of (NS+,NS+) and (R+,R+) fields are the same as Type II. 
Using these and other
properties it is possible to derive an expression for the effective gravity
action \cite{Klebanov:a} that, split in the NSNS and RR contributions,
reads:
\beqs
S_{\rm NSNS} = \int \dd^{10}x\;\sqrt{-g}\ee^{-2\Phi}\Big(R \nonumber\\
-\frac{1}{12} |\dd B|^2 + 4 |\dd\Phi|^2 - \frac{1}{2}|\dd T|^2 - V(T) \Big) 
\eeqs
\beqs
S_{\rm RR} = \int \dd^{10}x\;\sqrt{-g}\left(f(T)|F_{p+2}|^2+\cdots\right)
\eeqs
The main novelties are coming from the tachyon couplings. In particular, 
there is
a potential $V(T)$ which is an even function of the tachyon field, 
as well as functions $f(T)$ multiplying RR terms, that can be
worked out perturbatively.

\section{A couple of examples}
Here we recall the basic ideas behind applications of the Type 0 construction,
with the help of two of the earlier examples provided by Klebanov and
Tseytlin in the papers \cite{Klebanov:1999yy,Klebanov:1999ch}.
As in the usual Type II case, the idea is to consider a setup of 
D-branes of the theory, work out the gauge theory living on their world-volume,
and eventually find the corresponding solution of the gravity theory in
order to make comparisons and, hopefully, predictions.
    
\subsection{YM theory in 4 dimensions}
Consider the stack of $N$ D3 branes of the same type in Type 0B string theory.
The configuration is stable and the low energy spectrum on their world-volume 
consists of $SU(N)$ gauge bosons plus 6 adjoint scalars in $3+1$ dimensions.

In \cite{Klebanov:1999yy} the authors find two approximate solutions of the
gravity equations of motions that involve nontrivial tachyon, dilaton, together
with RR 4-form, and whose asymptotics are given in the UV and IR
regions\footnote{The identifications of these regimes follows from the
usual identification of the radial coordinate transversal to the stack, with 
energy scale of the gauge theory.}.

The metric approaches in both cases AdS$_5\times $S$^5$ (with different radii) 
and the behavior of the other fields is as follows.
The first one displays a vanishing 't Hooft coupling and a tachyon condensate
with $<T>\sim -1$, while the second represents an IR conformal point at infinite
coupling --- in fact the dilaton blows up while the tachyon condensate is
$<T>\sim 0$.

Interestingly, the asymptotic freedom is reproduced. It should 
be noted however that in this situation $\alpha'$ corrections become important, 
while 
in the IR one has to worry about string loop corrections. Moreover, connecting
the two asymptotic solutions is still an open question.
 
\subsection{Nonsupersymmetric CFT}
Another interesting problem that has been studied is the construction of gauge
theories that are conformal though nonsupersymmetric and their dual
gravitational description. Let us illustrate the construction in
\cite{Klebanov:1999ch}. In Type 0B theory one starts with a stack of $N$ D3 branes
of one type
(say, electric) and $N$ of the other (say, magnetic). 
This
configuration is argued to be stable in the large $N$ limit\footnote{At finite
N branes of different types repel each other because of fermionic degrees of
freedom in their world-volume.}.
The theory living on the world-volume of this stack is a $SU(N)\times
SU(N)$ gauge theory in $3+1$ dimensions, whose field content comprises, in addition
to gauge bosons, 6 adjoint scalars for each $SU(N)$ factor, 4 bifundamental fermions
in the $({\bf N},{\bf \bar N})$ and their conjugates.
The dual gravity solution found in \cite{Klebanov:1999ch} is again AdS$_5\times$S$^5$
with constant dilaton and vanishing tachyon. 

String loop suppression requires large $N$ as usual, while tachyon stability 
translates in a condition on the 't Hooft coupling, namely\\ $\lambda=g^2_{YM}
N\lta 100$.  Then the dual theory should be a CFT in the large $N$  limit, for not
very large coupling. 

The gravity solution resembles very much the Type IIB familiar case. In fact, it was
pointed out in \cite{Nekrasov:1999mn} that the gauge theory belongs to the
class of orbifold theories
of  ${\cal N}=4$ SYM, where the $Z_2$ projection belongs to the center of the
$R$-symmetry group. So it is exactly conformal in the large $N$ limit. This
information can be used to reverse the argument, that is, via AdS/CFT the gauge
theory may be predictive on the string theory (or gravity) side.
In fact the stability of the CFT at weak coupling implies that Type 0B theory on 
AdS$_5\times$S$^5$ should be stable for sufficiently small radius. In
\cite{Klebanov:1999um}
it is suggested that the instability of the background at large radius translates
in a phase transition occurring in the large $N$ CFT at strong coupling, with
anomalous dimension of the operator dual to the tachyon field developing a
singularity at a critical value $\lambda_c$.

\section{Non-critical Type 0}

In this section we expose some results obtained in \cite{Ferretti:a,Ferretti:b}
concerning the extension of Type 0 string theory in non-critical dimensions,
together with a proposal for the effective action and the discussion of
some solutions.
 
The original proposal in \cite{Polyakov:a} was to consider a string theory in 
dimension $d<10$. Even if a microscopic description of string theory out of 
criticality is still far, there are at least indications for a possible 
extension of Type 0 theories in non-critical dimensions:
\begin{itemize}
\item
a diagonal partition function, whose modular invariance doesn't rely 
on $d=10$ (as opposed to Type II theories)
\item
the tachyon should condense, providing an effective central charge 
$c_{\rm eff}\sim V(\langle T\rangle)$. It doesn't seem unnatural to shift  
$c_{\rm eff}$ by the central charge deficit $(10-d)$ 
\end{itemize}
$c_{\rm eff}$ provides a tree level cosmological constant in the low energy
theory,
and this agrees with the expectation that the inconsistencies possibly arise
only in flat background.

We work at the level of the effective gravity action. In $d<10$ one
should guess the field content of the theory and write down the relative
action. In \cite{Ferretti:b} we assume the NSNS sector (gravity + tachyon) 
is always 
present and the RR sector is worked out on group theory grounds, considering
tensor products of $SO(d-2)$ spinors. In five dimensions for instance, 
${\bf 2} \times {\bf 2}={\bf 1}+{\bf 3}$, and one is led to include
a scalar potential $A$ and a vector $A_{\mu}$. We further assume the existence
of a $4$-form potential to accommodate the would be D3-brane: This really 
amounts to consider massive gravity.    

With these assumptions, the equations of motion following from the
relative action
have interesting solutions, whose interpretation may give some insight
into their field theory duals, and eventually provide hints
in favor of the consistency of either critical or even non-critical
Type 0 string theory.
 
The relevant piece of the $d$-dimensional action, for the ansatz that 
we consider is
\footnote{Uppercase indices run from $1$ to $d$, while Greek and Latin 
indices run from $1$ to $p+2$ and from $p+3$ to $d$ respectively.}
\beqs
S=\int\dd^dx\sqrt{-g}\Bigg\{R\nonumber\\
 - \frac{1}{2} (\partial_{M}\Phi)^2 
- \frac{1}{2} (\partial_{M}T)^2 
- V(T)\ee^{\sqrt{\frac{2}{d-2}}\Phi} \nonumber \\ 
- \frac{f(T)}{2(p+2)!}\ee^{\frac{1}{2}\sqrt{\frac{2}{d-2}}(d-2p-4)\Phi} 
\Big(F_{M_1 \cdots M_{p+2}}\Big)^2  
\Bigg\},\nonumber
\eeqs
where $V(T)=-10+d-\frac{d-2}{8}T^2+\cdots$ is the tachyon potential, including 
central charge deficit.
  
One wants to find solutions of the equations of motion following from the
action above and
give them a dual interpretation. Consider an ansatz in which nonzero fields are
the metric, constant dilaton ($\Phi_0$) and tachyon 
($T_0$), and a RR ($p+2$)-form field strength: 
\beqs
R_{\mu\nu\rho\lambda}=-\frac{1}{R^2_0}\left(g_{\mu\rho}g_{\nu\lambda}
-g_{\mu\lambda}g_{\nu\rho} \right)\nonumber\\
R_{ijkl}=+\frac{1}{L^2_0}\left(g_{ik}g_{jl}-g_{il}g_{jk} \right) \\
F_{\mu_1 \cdots \mu_{p+2}} = F_0 \sqrt{-g_{(p+2)}} 
\epsilon_{\mu_1 \cdots \mu_{p+2}}~.\nonumber 
\eeqs
Then the equations of motion become a set of algebraic
equations. The tachyon VEV is determined implicitly by the following equation
\beqs
\frac{f'(T_0)}{f(T_0)} &=& \frac{1}{2}(d-2p-4)~ \frac{V'(T_0)}{V(T_0)}~.
\label{tacvev}
\eeqs
Now, without a precise knowledge of the functions $f(T)$ and $V(T)$ one
cannot infer whether it admits solutions. 
One should really assume it has, and extract some information.
The remaining equations fix the value
of the radii of the two maximally symmetric spaces and that of the dilaton.
It turns out that such solution is a 
AdS$_{p+2} \times$S$^{d-p-2}$ space, with tachyon
VEV, and fixed 't Hooft coupling $\lambda=e^{\Phi_0}N$.\footnote
{$N$ is the number of branes, which has to be evaluated in the string frame.}

Notice that string loop corrections are suppressed in the large $N$ limit, while
$\alpha'$-corrections are important because the curvature is $O(1)$.

By the AdS/CFT correspondence the field theory dual of this solution
should be at a conformal point. However it is difficult to make contact
with perturbative field theory because $\lambda\sim O(1)$. Nevertheless
one can still get additional information.
It is in fact possible to count the number of degrees of freedom  
that should live in the theory dual to this background \cite{Ferretti:a}.

Consider a thermal deformation of the solution. Identifying the Hawking  
temperature with the finite temperature of the field theory one can
compute the entropy. By either computing the free energy from the Euclidean
action, or computing the area of the horizon,
it turns out that it has the following behavior
\beqs
S\sim N^2V_pT_H^p 
\eeqs
for any value of $p$. This is an indication that the dual field theory should
have $N^2$ degrees of freedom, and could be YM theory in some non Gaussian 
limit. Notice
the different scaling power in the analogous relation one gets from evaluating
the entropies of black M2 and M5 branes ($3/2$ and 3 respectively).

Notice that this kind of approach is potentially 
predictive. Consider in fact the formula for scaling dimensions of
dual operators\footnote{These results are obtained in the case $d=p+2$.}
\beqs
\Delta &=& \frac{(d-1)+\sqrt{(d-1)+4m^2R_0^2}}{2}\\
m^2R_0^2 &=& d(d-1)\, \Bigg(1 + \frac{\tau}{2} 
\pm \frac{1}{2}\sqrt{\tau^2+(2d-4)\frac{V'(T_0)^2}{V(T_0)^2}}\Bigg)\nonumber
\eeqs
\beqs
\qquad {\rm with} \qquad 
\tau = d\frac{V'(T_0)^2}{V(T_0)^2} -\frac{2}{d}\frac{f''(T_0)}{f(T_0)}
-\frac{V''(T_0)}{V(T_0)}-1~.\nonumber
\eeqs

First, note that the tachyon VEV $T_0$ behaves as a ``bare'' quantity: 
it does not
enter in determining physical quantities. It is very much as in the
renormalization group. One can do a field redefinition, this will shift
$T_0$, without affecting $\lambda$, $R_0$, $\Delta$.

Then, the scaling dimensions depend on a {\it finite} number of 
parameters and with a good guess
on the field theory side, one in principle should be able to predict an 
infinite tower of dimensions from KK analysis. 

\section{Holographic RG Flows}

The extension of the AdS/CFT correspondence to theories away from their conformal
limit leads naturally to the issue of giving a dual description of 
RG flows. This {\it holographic} description has recently attracted much attention.
Here we will focus on a Type 0 application.

The construction of gravity solutions describing RG flows is quite general.  Given
1) an effective potential for the scalars and 2) the existence of AdS solution (at
extrema of the potential), the procedure is in principle straightforward.
In fact the equations of motion for the coupled scalars plus gravity system provide
the RG equations for the couplings in the dual theory, once an appropriate ansatz
for the metric is plugged in as a physical input.
This setup can be achieved considering consistent truncations
of Type IIB or 11d supergravities (for instance gauged supergravities) or other
KK compactifications emerging for example as compactifications on manifolds
which are near horizon limit of branes at conical singularities.
Finally, such a framework can be provided by a simple gravity theory, such as the
Type 0 gravity. 

In this case the effective action is:
\beq
S=\int \dd^dx \sqrt{-g} \left\{R - \frac{1}{2}(\de\phi_i)^2-
{\cal V}(\phi_i)\right\}~.
\eeq
where $\phi_i=(\Phi,T)$.
With the following ansatz
\beqs
\dd s^2=\dd y^2+A^2(y)\dd\vec{x}^2\nonumber\\
\phi_i=\phi_i(y)
\eeqs
the equations of motions read
\beqs
\ddot\phi_i+\gamma\dot\phi_i=\de_i{\cal V}\nonumber\\
\frac{\ddot A}{A}+{(d-2)}\left(\frac{\dot A}{A}\right)^2=\frac{1}{d-2}{\cal
V}\label{particle}\\
\dot\gamma=-\frac{d-1}{2(d-2)}(\dot\phi_i)^2\nonumber
\eeqs 
where $\gamma=(d-1)\frac{\dd}{\dd y}\log A$. 

Now, suppose that (\ref{tacvev}) has more than one solution, say $T_1, T_2$ 
at least. This
means that exist two AdS solutions, with fixed values for scalars, and
radii $R_2>R_1$, say.
Then, it can be shown (see \cite{Petrini:2000qa} for instance) that there are
interpolating solutions between the previous two. The explanation of this fact
is roughly as follows. The set of equations (\ref{particle}) can be interpreted 
as describing a point-like particle rolling in a potential $-{\cal V}$, subject
to an effective friction force encoded in $\gamma$. In fact, given that $\gamma$
is monotonically decreasing and 
\beq
\gamma \to \frac{d-1}{R_2}>0 
\eeq  
as $y\to \infty$, it follows that it is strictly positive along the trajectory.
The solutions
we are looking for are then those starting at a minimum of $\cal V$ for $y\to
-\infty$ (IR) and ending at maximum for $y\to \infty$ (UV). 

Some universal information can be read from the local behavior of these solutions.
Consider for instance the UV fixed point. First it can be identified a 
coordinate which in the field theory can be consistently
interpreted as the energy scale and parameterizes the interpolating solution.
Such a coordinate can be chosen to be\footnote{Different definitions
can be given, which reduce locally to the same one.}
\beq
U=\frac{A^2}{\dot{A}}~.
\eeq
The linearized equation for the fluctuations eigenvectors $\delta\tilde\phi_i(U)$ 
has the solutions
\beq
\delta\tilde\phi_i=A_i U^{-(d-1-\Delta_i)}+B_i U^{-\Delta_i}
\eeq
where $\Delta_i$ were given in the previous section. One can then extract the leading
$\beta$-functions for the running couplings\footnote{Unless $\Delta_i=d-1$, which
means that there is a VEV for some operator in the dual field theory
\cite{Petrini:2000qa}.}:
\beq
\beta_i(g_i)=U\frac{\dd}{\dd U}g_i=(\Delta_i-(d-1))(g_i-g_i*)+\cdots
\eeq

\section{R\'esum\'e and Perspectives}
We have illustrated how the Type 0 approach can be used to apply the AdS/CFT 
correspondence to nonsupersymmetric cases. This method offers some advantages
compared to others approaches, as the finite temperature one for instance. 
Though,
it has some limitations as well. We conclude summarizing some problems 
\begin{itemize}
\item
${\cal N}=0\to$ difficult to check\footnote{This is common to all nonsupersymmetric
approaches.}
\item
there is no {\it proof} of tachyon condensation 
\item
$\alpha'$ corrections are usually important
\item
lack of a microscopic stringy description in $d<10$
\end{itemize}
and some interesting features
\begin{itemize}
\item
cutoff independent results
\item
confining and asymptotically free solutions
\item
no mixing with higher dimensional physics
\item
``conformal'' solutions 
\end{itemize}
In particular the last point is potentially interesting. These solutions could be 
just an artifact of the approximation. However they can also indicate some novel 
field theories or (strongly coupled) conformal fixed points in YM theories as
suggested in \cite{Polyakov:a}. 

We conclude noticing that, some control on the field
theory may shed light on Type 0 string theory and/or on
non-critical string theory via AdS/CFT correspondence. In particular the latest 
developments in the context of holographic RG flows could provide an appropriate
framework for studying which constraints the dual field 
imposes on the gravity theory.

\begin{acknowledgments}
I wish to thank the organizers of the conference for hospitality. 
This work was supported in part by European Union TMR program CT960045.
\end{acknowledgments}

\end{document}